\documentclass[11pt]{article}
\usepackage{moriond}
\usepackage{units}
\usepackage{xspace}
\usepackage{amsmath}

\newcommand{\inmath}[1]{\ensuremath{#1}\xspace}

\newcommand{\GeV}[1]{{\unit[#1]{GeV}}}
\newcommand{\TeV}[1]{{\unit[#1]{TeV}}}
\newcommand{\ipb}[1]{\inmath{\unit[#1]{pb^{-1}}}} 
\newcommand{\ifb}[1]{\inmath{\unit[#1]{fb^{-1}}}} 
\newcommand{\Fig}[1]{Figure~\ref{fig:#1}\xspace}
\newcommand{\met  }{\inmath{E_T^\textnormal{miss}}}
\newcommand{\meff }{\inmath{m_\text{eff}}}
\newcommand{\pt   }{\inmath{p_T}} 

\newcommand{\seventev}{\inmath{\sqrt{s} = \TeV{7}}}
\newcommand{\eighttev}{\inmath{\sqrt{s} = \TeV{8}}}
\newcommand{\percent}[1]{{#1}\,\%}
\newcommand{\gluino       }{\inmath{{\skew{4}{\widetilde}{g}}}}
\newcommand{\squark       }{\inmath{{\skew{4}{\widetilde}{q}}}}
\newcommand{\gravitino    }{\inmath{{\skew{6}{\widetilde}{G}}}}
\newcommand{\neutralinon  }[1]{\inmath{\widetilde{\chi}_{#1}^0}}
\newcommand{\stau         }{\inmath{\widetilde{\tau}}}
\newcommand{\onehalf      }{\ensuremath{{\nicefrac12}}}
\newcommand{\mzero        }{\inmath{{m_0}}}
\newcommand{\moh          }{\inmath{{m_\onehalf}}}
\newcommand{\ttbar}{\inmath{t\skew{3}{\bar}{t}}}
\newcommand{\Wjets}{\inmath{W\!+\text{jets}}}

\newcommand{\Photo}{}

\begin{document}

\vspace*{4cm}
\title{SUSY Searches for Inclusive Squark and Gluino Production at the LHC}

\author{ A. MANN \\
On behalf of the ATLAS and CMS Collaborations}

\address{
Ludwig-Maximilians-Universit\"at M\"unchen,
Fakult\"at f\"ur Physik, \\
Am Coulombwall 1,
85748 Garching, Germany}

\maketitle\abstracts{
A selection of recent searches for Supersymmetry 
conducted by the ATLAS and CMS collaborations is presented,
summarizing the relevant results for inclusive squark and gluino production at the LHC.
No indication for the existence of Supersymmetry 
in terms of an excess above the Standard Model background is seen by any analysis,
therefore limits are reported on parameters of the various models used to interpret the findings.}

\section{Introduction}

After a long and successful data-taking period from 2010 through 2012,
the first run of the Large Hadron Collider (LHC) at CERN has come to an end.
In 2012, both the ATLAS and the CMS collaboration have taken
about \ifb{22} of data at \eighttev,
about four times the data available from 2010 and 2011
at a lower center-of-mass energy of \seventev.

Already with the first proton-proton collision data from 2010 of about \ipb{35},
the searches for inclusive squark and gluino production 
proved to be very successful in terms of mass reach,
yielding exclusion limits that rapidly surpassed those of earlier experiments.
Since then, the analyses have evolved and diversified, 
pushing the limits on the squark, gluino and lightest gaugino masses or mass parameters to higher and higher values,
aiming at more complex topologies and giving additional interpretations in other supersymmetric models.

In these proceedings, the focus will be on strong production of the 
scalar partners of first and second generation quarks
and the fermionic partners of the gluons.
For a given supersymmetric mass scale, 
strong production has the largest production cross sections,
thus allowing to target heavy initial supersymmetric particles,
signatures with small branching ratios and long decay chains.
Apart from one exception all analyses presented below assume $R$-parity conservation,
which makes the lightest supersymmetric particle (LSP) stable.
Missing transverse momentum (\met) is therefore an important part of the searched for signature,
the other being jets with high transverse momentum 
produced 
in the decay chains starting from the heavy pair-produced squarks and gluinos.
%
Only results by the ATLAS and the CMS collaborations are considered
which have been made public on or before the 11th March 2013,
the date at which the corresponding talk has been given.
None of the analyses has seen any significant excess of events
above the expected number from the Standard Model predictions,
thus all analyses interpret their findings by reporting limits on supersymmetric particle masses or model parameters.

\section{Searches with the Full 2012 Dataset}

The ATLAS collaboration has made public results from two searches for Supersymmetry 
with inclusive squark (\squark) and gluino (\gluino) production which make use of the full 2012 dataset.

The first search~\cite{ATLAS-CONF-2013-026} targets final states with exactly one, 
or two or more hadronically decaying tau leptons, together with jets and \met.
The motivation for this search comes from models with gauge-mediated supersymmetry breaking (GMSB)
where the scalar partner of the tau, the stau \stau, is the next-to-lightest supersymmetric particle (NLSP),
leading to final states with many taus.
As no excess is seen,
the results are interpreted in terms of limits at \percent{95} confidence level in three different models.
The limits in the mSUGRA/CMSSM framework 
are shown in \Fig{ATLAS-CONF-2013-026_05},
where this analysis is in particular powerful in the low \mzero region.
The model parameters for the mSUGRA/CMSSM model have been chosen 
such that in a large part of the plane 
a \GeV{126}~Higgs can be accommodated.
Not shown is the limit plot for the minimal GMSB model 
in the plane spanned by the parameters~$\Lambda$, 
the SUSY breaking mass scale felt by the low-energy sector,
and~$\tan \beta$, the ratio of the MSSM Higgs vacuum expectation values.
Here, the previous limits from 2011 data can be extended considerably,
reaching now up to \TeV{70} in~$\Lambda$ for high $\tan \beta > 50$ 
(corresponding to $m(\gluino)$ of about $\TeV{1.5}$),
and excluding values of~$\Lambda$ below \TeV{54} 
($m(\gluino) \sim \TeV{1.2}$) 
independent of $\tan \beta$. 
There is also a new interpretation in a model of natural gauge mediation,
where independently of the \stau mass, gluino masses up to \TeV{1.14} can be excluded, 
provided the \stau is the NLSP. 

The second analysis \cite{ATLAS-CONF-2013-007} 
looks at final states with same-sign lepton pairs,
together with $3$ or more jets and \met. 
The motivation for this analysis is that the pair-produced gluinos have almost the same probability 
to give pairs of leptons with same charge or with opposite charge.
Requiring a pair of leptons with same sign suppresses the background coming from Standard Model processes considerably,
giving a very clean and powerful signature to look for new physics processes.
%
%
The limits which are set 
in the same mSUGRA/CMSSM model with Higgs-compatible parameters used for the tau analysis 
are shown in \Fig{ATLAS-CONF-2013-007_07},
excluding a large region of the mSUGRA/CMSSM parameter space. 
In addition, interpretations in a large number of simplified models are given,
demonstrating the power and versatility of this search channel.

\begin{figure}
  \begin{minipage}{0.5\linewidth}
    \centering
    \includegraphics[width=0.9\linewidth]{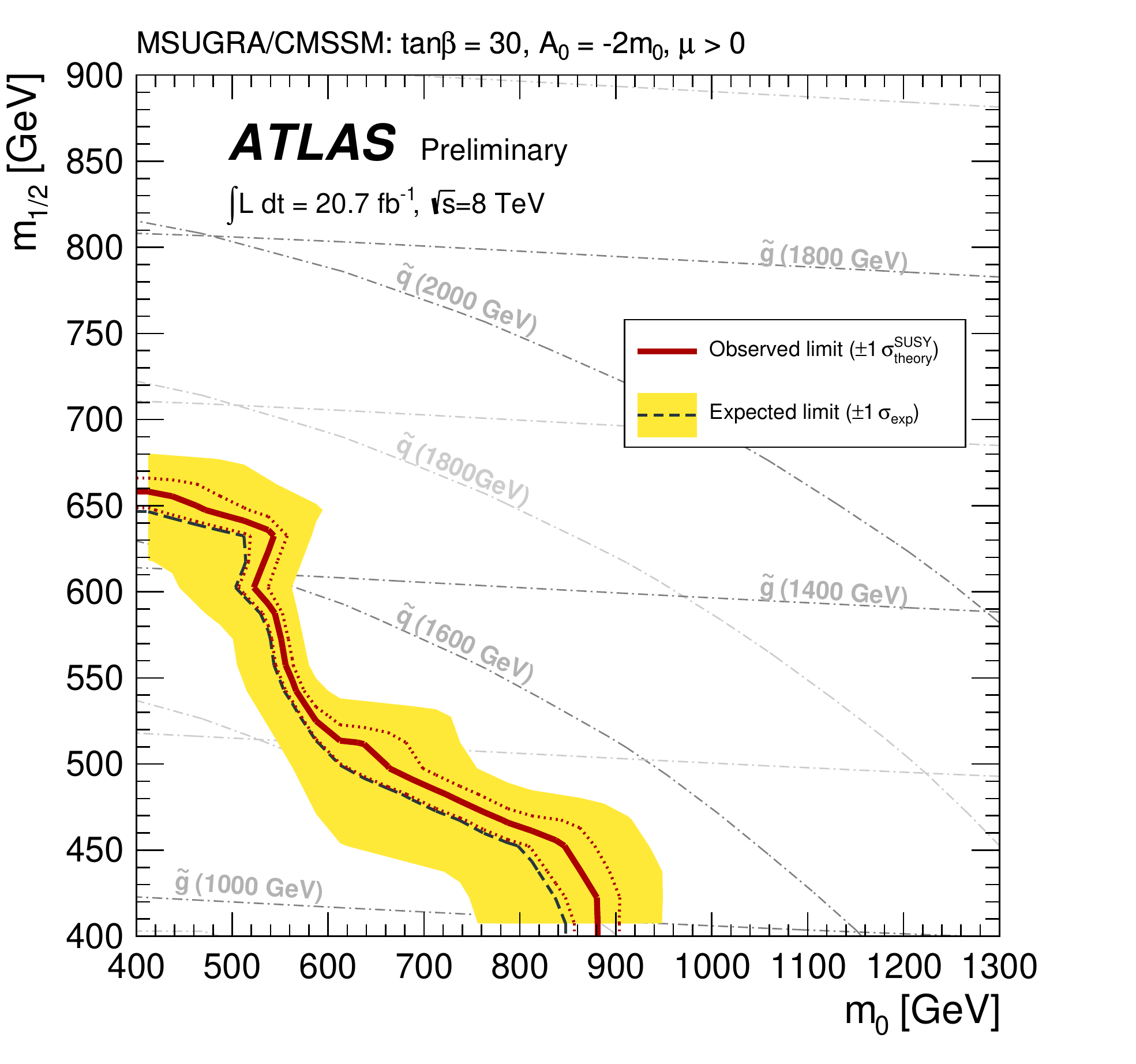}
  \end{minipage}
  \hfill
  \begin{minipage}{0.5\linewidth}
    \includegraphics[width=0.9\linewidth]{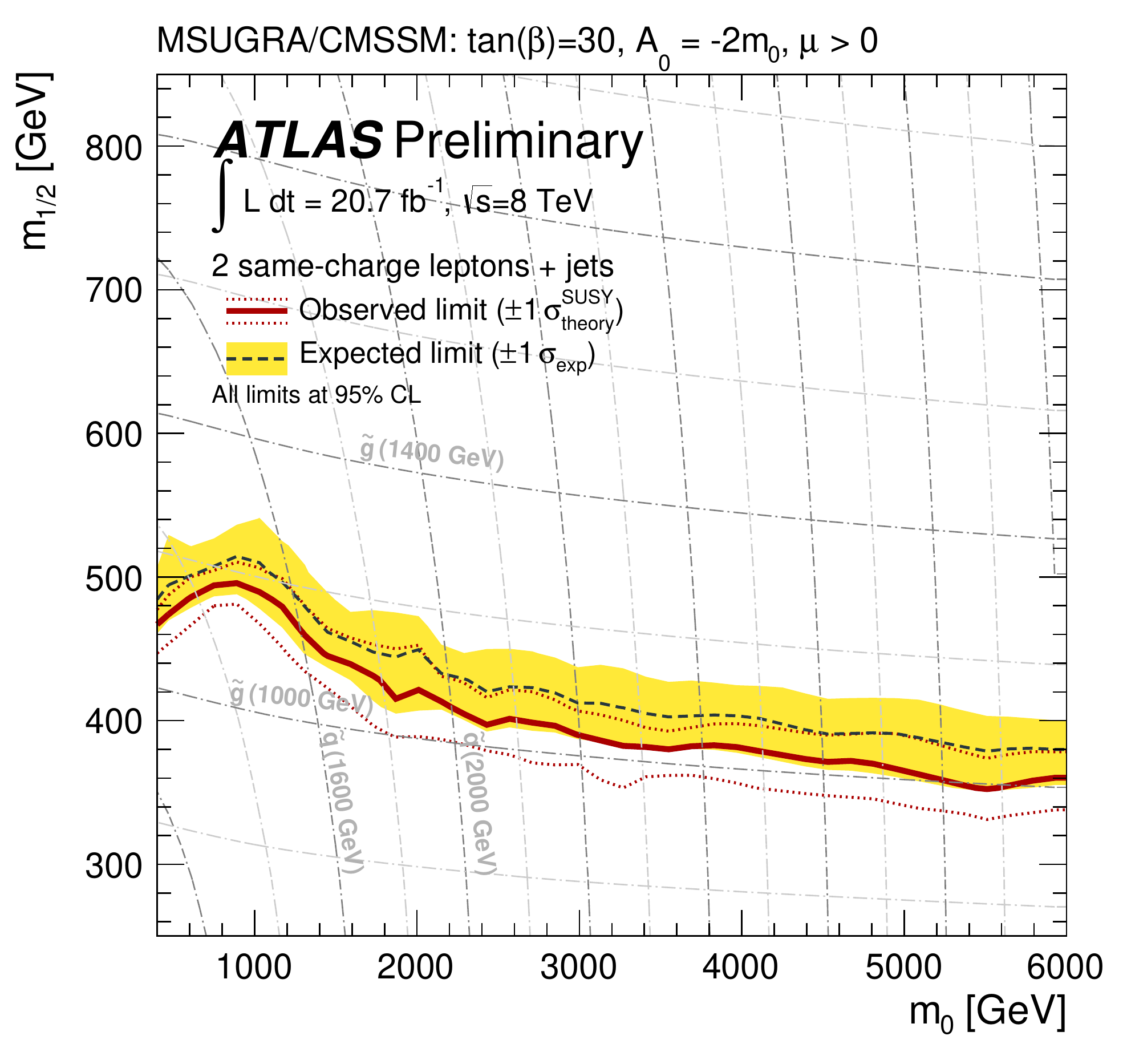}
  \end{minipage}
  \caption{
    Expected and observed limits from the two ATLAS analyses making use of the full 2012 dataset. 
    The plots show the interpretation of the 
      tau + jets + \met~results (left, \protect\cite{ATLAS-CONF-2013-026})
    and 
      same-sign leptons + jets + \met~results  (right, \protect\cite{ATLAS-CONF-2013-007})
    in the mSUGRA/CMSSM framework.
    (Note the very different range on the horizontal axis.)
  }
  \label{fig:ATLAS-CONF-2013-026_05}
  \label{fig:ATLAS-CONF-2013-007_07}
\end{figure}

\section{Searches with a Subset of 2012 Data}

CMS has made public a study using \ifb{9.2} from the 2012 dataset~\cite{CMS-PAS-SUS-12-027},
in which models with $R$-parity violation (RPV) are considered.
As here the LSP is no longer stable, 
high particle multiplicities are characteristic for supersymmetric decays rather than large \met.
Events with $3$ or more isolated leptons are selected and classified with respect to
the number of opposite-sign same-flavour (OSSF) lepton pairs,
the invariant mass of the OSSF lepton pairs if present,
the presence of jets from bottom quarks, 
and the presence of a hadronically decaying tau.
Together with a binning in $S_T$, 
defined as the scalar sum of \met and the transverse momentum (\pt) of jets and leptons,
this gives in total $240$~event classes,
which are then combined to derive exclusion limits.
Numerous models with different topologies are considered,
probing all three RPV Yukawa couplings individually.
In the RPV squark-gluino models,
lower limits are set on $m(\gluino)$ and $m(\squark)$ of the order of $1$ to \TeV{2}.
In a CMSSM with RPV, $\moh < \TeV{1.15}$ is excluded. 
An update with the full 2012 dataset,
limited to interpretations with light scalar top quarks 
and non-zero leptonic or semileptonic RPV Yukawa couplings%
, has been made public recently~\cite{CMS-PAS-SUS-13-003}. 

Both ATLAS and CMS have analyses motivated from models of general gauge-mediation (GGM)
with the lightest neutralino $\neutralinon1$ being the NLSP \cite{ATLAS-CONF-2012-152,CMS-PAS-SUS-12-018}.
ATLAS considers a higgsino-like $\neutralinon1$ which decays predominantly to a $Z$~boson and the gravitino LSP. 
Thus, events with jets, \met and same-flavour leptons with an invariant mass compatible with a $Z$~boson are selected.
Limits are set in the plane spanned by the higgsino mass parameter $\mu$, 
which is roughly equal to the neutralino mass,
and the gluino mass.
Now using \ifb{5.8} of data,
the limits for $\tan\beta = 1.5$ improve on the 2011 results from the same analysis considerably. 
At a larger value of $\tan \beta = 30$, the decay $\neutralinon{1}\to h\gravitino$ also becomes relevant.
The excluded area for this value of $\tan \beta$, however, shown in \Fig{ATLAS-CONF-2012-152_fig_05},
is barely smaller than that at $\tan \beta = 1.5$.
CMS on the other hand considers a bino- or wino-like NLSP, 
such that the decay to photon plus gravitino or $Z$~plus gravitino, respectively, dominates.
Two separate analyses look for events with $1$~photon and $2$~or more jets, 
or $2$~photons and $1$~or more jet, each with a loose selection on \met.
With respect to their exclusion limits in the squark and gluino mass plane,
the two analyses are comparable,
the one-photon channel giving stronger limits in the scenario with a wino-like neutralino 
where less photons are expected due to the competing $\neutralinon{1} \to Z\gravitino$ decays becoming more probable.
From \ifb{4} of data, $m(\gluino) < \TeV{1.1}$ and $m(\squark) < \TeV{1.25}$ for a bino-like 
and $m(\gluino) < \GeV{750}$ and $m(\squark) < \GeV{850}$ for a wino-like neutralino with $m(\neutralinon1)=\GeV{375}$ are excluded. 

\begin{figure}
  \begin{minipage}{0.5\linewidth}
    \centering
    \includegraphics[width=0.95\linewidth]{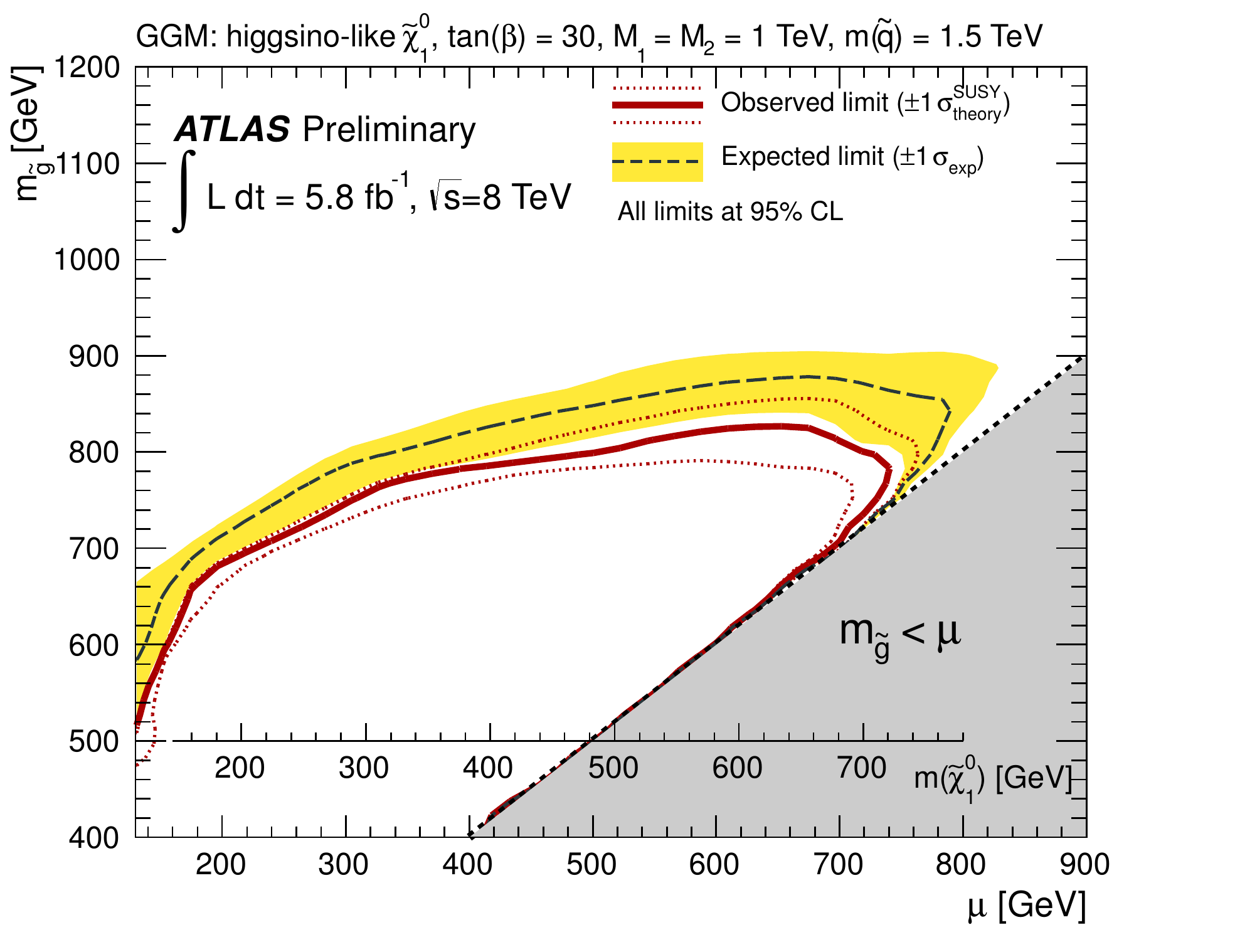}
  \end{minipage}
  \hfill
  \begin{minipage}{0.5\linewidth}
    \centering
    \includegraphics[width=0.9\linewidth]{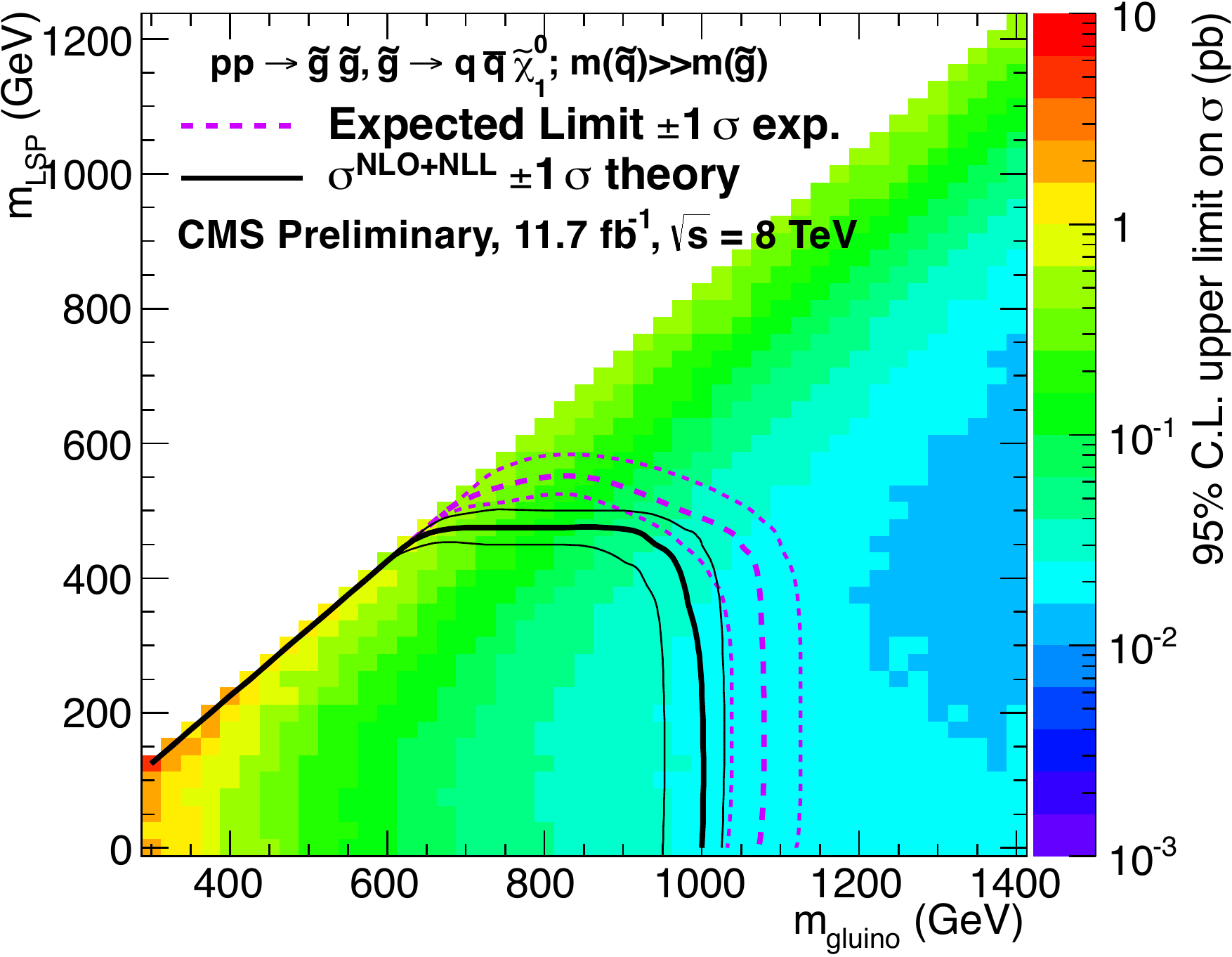}
  \end{minipage}
  \caption{
    Left:  Exclusion limits on the gluino mass and $\mu$ parameters for the GGM models with $c\tau_\textnormal{NLSP} < \unit[0.1]{mm}$~\protect\cite{ATLAS-CONF-2012-152}. 
    Right: Upper limit on the cross section at \percent{95} confidence level as function of $m(\gluino)$ and $m(\neutralinon{1})$. 
      The solid black line indicates the observed excluded region \protect\cite{CMS-PAS-SUS-12-028}.
  }
  \label{fig:ATLAS-CONF-2012-152_fig_05}
  \label{fig:CMS-PAS-SUS-12-028_T1}
\end{figure}

The zero-lepton analyses are the latest in a series of inclusive searches for squark and gluino production.
In ATLAS, this channel is split into two sub-channels with minimum jet multiplicities of $2$ to $6$ and $6$ to $9$, respectively.
The most recent publications for both sub-channels use \ifb{5.8} of 2012 data.
High jet multiplicities are expected from decays of gluino pairs via off-shell stops,
therefore this topology is targeted in the analysis of events with at least~$6$ to at least~$9$ jets~\cite{ATLAS-CONF-2012-103}. 
The search with at least~$2$ to at least~$6$ jets~\cite{ATLAS-CONF-2012-109} targets 
final states
where each initial squark yields one jet plus \met, 
and each initial gluino yields two jets plus \met. 
Selection cuts on the ratio of the effective mass \meff, 
defined as the scalar sum of \met and the \pt of the jets,
the ratio of $\met/\meff$,
and the minimum angle between jets and \met in the transverse plane 
are used to suppress the dominant multi-jet background.
Events with electrons or muons are rejected. 
The limits on $m(\squark)$ and $m(\gluino)$
are found to be similar for the mSUGRA/CMSSM implementation  
and a simplified squark-gluino-neutralino MSSM model:
Equal mass light-flavor squarks and gluinos with masses below \TeV{1.5} can be excluded.
(Note that the values of the parameters for the mSUGRA/CMSSM grid used here
differ from those used in the interpretation of the full-search datasets described above.)
%
%
The corresponding zero-lepton analysis from CMS~\cite{CMS-PAS-SUS-12-028}
has a different approach to suppress the multi-jet background.
It makes use of a kinematic variable computed from the jets in each event, denoted by $\alpha_T$~\cite{CMS-PAS-SUS-08-005}, 
to select only events with significant genuine \met.
In addition, selected events must have at least 2 jets, no isolated electrons, muons nor photons,
and $H_T > \GeV{275}$, where $H_T$ is the defined as the scalar sum of the transverse energies of the jets.
The events are then categorised by the number of jets ($2$ to $3$ or $4$ or more) and the number of $b$-tagged jets 
and binned in $H_T$.
Standard Model backgrounds are estimated from data control regions,
which are mapped to the signal regions using transfer factors obtained from Monte Carlo.
Systematic uncertainties on these transfer factors are derived from a set of closure tests.
Simplified models are used to interpret the findings 
in terms of limits on squark or gluino and neutralino masses.
With \ifb{11.7} of integrated luminosity,
in the models with gluino or squark pair production 
this search is reported to be sensitive 
up to $m(\squark,\gluino) \approx \GeV{950}$, $m(\neutralinon1) \approx \GeV{450}$ 
or $m(\squark,\gluino) \approx \GeV{775}$, $m(\neutralinon1) \approx \GeV{325}$,
respectively.

Finally, there is also an ATLAS result with 2012 data 
for the one-lepton channel with high jet multiplicity \cite{ATLAS-CONF-2012-104}, using \ifb{5.8}.
The results are interpreted 
in the same mSUGRA/CMSSM plane as used in the ATLAS zero-lepton analysis described above.
Events with exactly one isolated electron or muon, 
$4$ or more jets and large $\met > \GeV{250}$ are selected. 
Additional selection cuts on the transverse mass of the lepton, \meff and $\met/\meff$ are applied,
after which \Wjets and \ttbar production remain as dominant background processes in the signal region.
The background reaching the signal region is estimated using a global fit.
For equal squark and gluino masses, values up to \TeV{1.24} are excluded in the mSUGRA/CMSSM interpretation,
for large \mzero the limit on the gluino mass is about \GeV{900}. 
CMS also made public a 2012 update of their one-lepton analysis 
with interpretations in models of direct top squark pair production.

Overall, there have been many interesting updates from both experiments,
demonstrating already now how much the constraints on supersymmetric models can be extended
using the full 2012 dataset.
None of the analyses has reported any significant excess above the Standard Model background,
but on the other hand many more important results will be made public in the course of 2013.
And, of course, the new data at higher center-of-mass energy anticipated for 2015 after the restart of the LHC is eagerly awaited.

\section*{References}

The publications listed below are freely available from the websites of the ATLAS and CMS experiments and the CERN Document Server (\url{http://cds.cern.ch}).


\end{document}